\documentclass[journal]{IEEEtran}

\ifCLASSINFOpdf
\else
   \usepackage[dvips]{graphicx}
\fi
\usepackage{url}

\usepackage{graphicx}
\usepackage{algorithmic, amsfonts}
\usepackage{amsmath, amssymb}
\usepackage{cite, array, multirow}
\usepackage{setspace}

\renewcommand{\baselinestretch}{0.88}

\begin{document}

\title{Complex Neural Spatial Filter: Enhancing Multi-channel Target Speech Separation in Complex Domain}

\author{Rongzhi Gu, Shi-Xiong Zhang, Yuexian Zou, and Dong Yu
\thanks{This research is partly supported by Shenzhen Science \& Technology Fundamental Research Programs (No: JCYJ20180507182908274 \& JSGG20191129105421211). }
\thanks{Rongzhi Gu and Yuexian Zou (corresponding author) are with the Advanced Data and Signal Processing laboratory (ADSPLab), School of Electric and Computer Science, Peking University Shenzhen Graduate School, Shenzhen, China. Yuexian Zou is also with Pengcheng Laboratory, Shenzhen, China. e-mail: zouyx@pku.edu.cn. }
\thanks{Shi-Xiong Zhang and Dong Yu are with Tencent AI Lab, Seattle, WA, USA.}}

\markboth{}
{Shell \MakeLowercase{\textit{et al.}}: Bare Demo of IEEEtran.cls for IEEE Journals}
\maketitle

\begin{abstract}
To date, mainstream target speech separation (TSS) approaches are formulated to estimate the complex ratio mask (cRM) of target speech in time-frequency domain under supervised deep learning framework. However, the existing deep models for estimating cRM are designed in the way that the real and imaginary parts of the cRM are separately modeled using real-valued training data pairs.
The research motivation of this study is to design a deep model that fully exploits the temporal-spectral-spatial information of multi-channel signals for estimating cRM directly and efficiently in complex domain.
As a result, a novel TSS network is designed consisting of two modules, a complex neural spatial filter (cNSF) and an MVDR. Essentially, cNSF is a cRM estimation model and an MVDR module is cascaded to the cNSF module to reduce the nonlinear speech distortions introduced by neural network. Specifically, to fit the cRM target, all input features of cNSF are reformulated into complex-valued representations following the supervised learning paradigm. Then, to achieve good hierarchical feature abstraction, a complex deep neural network (cDNN) is delicately designed with U-Net structure. Experiments conducted on simulated multi-channel speech data demonstrate the proposed cNSF outperforms the baseline NSF by 12.1\% scale-invariant signal-to-distortion ratio and 33.1\% word error rate.

\end{abstract}


\IEEEpeerreviewmaketitle

\section{Introduction}

\IEEEPARstart{T}{arget} speech separation (TSS) aims to recover the target speech that is corrupted by interfering speech, noise and reverberation. TSS is one of the most important yet challenging tasks for robust automatic speech recognition (ASR) \cite{chen2018multi,yoshioka2018multi,wang2018supervised}.

With the entry into deep learning era, significant improvements have been witnessed in TSS performance \cite{wang2018voicefilter, ephrat2018looking,vzmolikova2019speakerbeam,xiao2019single}.
Based on the time-frequency (T-F) sparsity assumption of speech signals \cite{watanabe2017new}, mainstream deep learning based TSS methods formulate TSS task as a supervised learning problem in T-F domain to learn a mask \cite{wang2018supervised}, which indicates whether the target speech dominates a T-F bin.
In the early study, the learning objective was to estimate a magnitude mask of the target speech from the mixture magnitude, leading to ratio mask (RM) based TSS methods \cite{yu2017permutation,hershey2016deep}. Essentially, these RM based methods could be viewed as working in the magnitude domain and ignoring the phase estimation. Later research showed that the reuse of the mixture phase to reconstruct the target speech becomes a limiting factor to the performance, which decreases the intelligibility of the estimated target speech \cite{le2019phasebook}. To tackle this problem, phase-sensitive mask (PSM) and complex ratio mask (cRM) were further proposed to consider the phase information \cite{wang2014training, takahashi2018phasenet, yin2020phasen,wang2018alternative}.
Specifically, cRM is defined as the division of the complex spectrogram of clean speech and its mixture \cite{wang2014training}:
\vspace{-0.1cm}
\begin{equation}
    M = \frac{S}{Y} = \frac{Y_r S_r + Y_i S_i }{|Y|^2} + j \frac{Y_r S_i - Y_i S_r }{|Y|^2} =M_r+jM_i
\label{eq:crm}
\end{equation}
where $M$ is cRM, $S$ denotes the complex spectrogram of target speech, $Y$ is the complex spectrogram of single-channel mixture signal or the reference channel of multi-channel mixture signal, $j=\sqrt{-1}$ is the imaginary unit, $|\cdot|$, $(\cdot)_r$ and $(\cdot)_i$ take the magnitude, real and imaginary parts of a complex-valued number, respectively. For presentation conciseness, the frame index and frequency index are omitted in Eq.\ref{eq:crm}. Clearly, from Eq. \ref{eq:crm}, the target spectrogram $S$ can be reconstructed from $Y$ perfectly if $M$ is correctly estimated. Therefore, the estimation of cRM becomes the key for TSS task. Recently, cRM based TSS methods have become the mainstream, where the real and imaginary parts of cRM are separately estimated.

The discussion above is only related to the single-channel TSS task. It is obvious that spatial information is an effective cue that can be exploited to further enhance the TSS performance. As a result, TSS with multiple microphones has drawn great attention recently.
Previous investigations on interaural phase difference (IPD) and its derived directional features showed that these features can provide discriminative cues for estimating the cRM of the target speech \cite{wang2018multi, lianwu2019multi,yoshioka2018multi,wang2018spatial, wang2019combining, chen2018multi}. Our previous work, neural spatial filter (NSF) \cite{gu2019neural}, leveraged the direction information of the target speech and formulates two directional features to indicate the dominance of acoustic components from the target direction, which guides the deep neural network (DNN) model to learn better masks. Very recently, to reduce the nonlinear speech distortions brought by DNN model, based on the classic minimum variance distortionless response (MVDR) \cite{xu2020neural} beamforming technique, Xu et. al. proposed to jointly train the DNN and MVDR beamformer to improve the robustness of MVDR coefficients estimation. Experimental results demonstrated that both the ASR accuracy and perceptual speech quality are improved.

There is also some exploring work related to cRM estimation for monaural speech enhancement (SE) task.
\cite{yin2020phasen} investigated to estimate the cRM using two-stream DNNs that process magnitude and phase respectively. Accordingly, $\{|Y|, |M|\}$ and $\{\angle Y, \angle M\}$ paired data were formed to supervise the training of these two networks. Since the distribution of phase does not have an explicit structure as magnitude, to promote the learning of phase, a communication module was designed to facilitate the information flow between the magnitude and phase stream.

As discussed above, existing cRM based methods model the magnitude and phase (or real and imaginary) of the cRM separately with two sets of real-valued training pairs.
This brings one question: is it a good approach to estimate the complex-valued cRM using two-stream real-valued neural network? In other words, is it possible to estimate the cRM directly using complex deep neural network (cDNN) \cite{trabelsi18deepcomplex}? This idea is essentially triggered by the cDNN models introduced recently to support phase-aware processing for SE task \cite{choi2018cunet, hu2020dccrn, pandey2019exploring}. Bearing this idea, our work focuses on the multi-channel TSS task and makes efforts to design a single-stream framework, which consists of a cRM estimation network (named as complex neural spatial filter, cNSF) and an MVDR module. cNSF works in complex domain, which is delicately designed to fully exploit the temporal-spectral-spatial information. Also, to suppress the nonlinear speech distortion introduced by DNN, an MVDR module is introduced into the TSS framework and jointly trained with cNSF.
More specifically, following the supervised learning paradigm, we reformulate the training pair to fit the cRM target, where all input features of cNSF are reformulated into complex-valued representations. Secondly, to support the processing in complex domain, a cDNN is carefully designed where all the network components and operations are in complex domain. To achieve hierarchical feature abstraction, the network is designed in the form of U-Net structure \cite{ronneberger2015unet} with dense CNN blocks \cite{huang2017densely} to strengthen feature capturing and bidirectional long short term memory (BLSTM) layers to enhance the sequential information modeling. Thirdly, to further reduce the nonlinear speech distortions and attenuate the interferences from the undesired direction, an MVDR neural beamformer is adopted and contributes to the training process.

The contributions of this work can be summarized as follows: 1) a multi-channel TSS framework is formulated in complex domain under supervised learning paradigm, which aims to estimate the cRM directly using a single-stream network with unified metric; 2) A novel cNSF-MVDR multi-channel TSS model is designed, which fully exploits the temporal-spectral-spatial information of the multi-channel signal; 3) To verify our proposed cNSF-MVDR model, extensive experiments have been conducted with simulated multi-channel speech data. Experimental results demonstrate the proposed cNSF-MVDR outperforms the baseline NSF by 12.1\% scale-invariant signal-to-distortion ratio (SI-SDR) and 33.1\%word error rate (WER).


\section{Proposed TSS Framework}
\label{sec:cnsf}

\begin{figure*}
\centerline{\includegraphics[width=15.5cm]{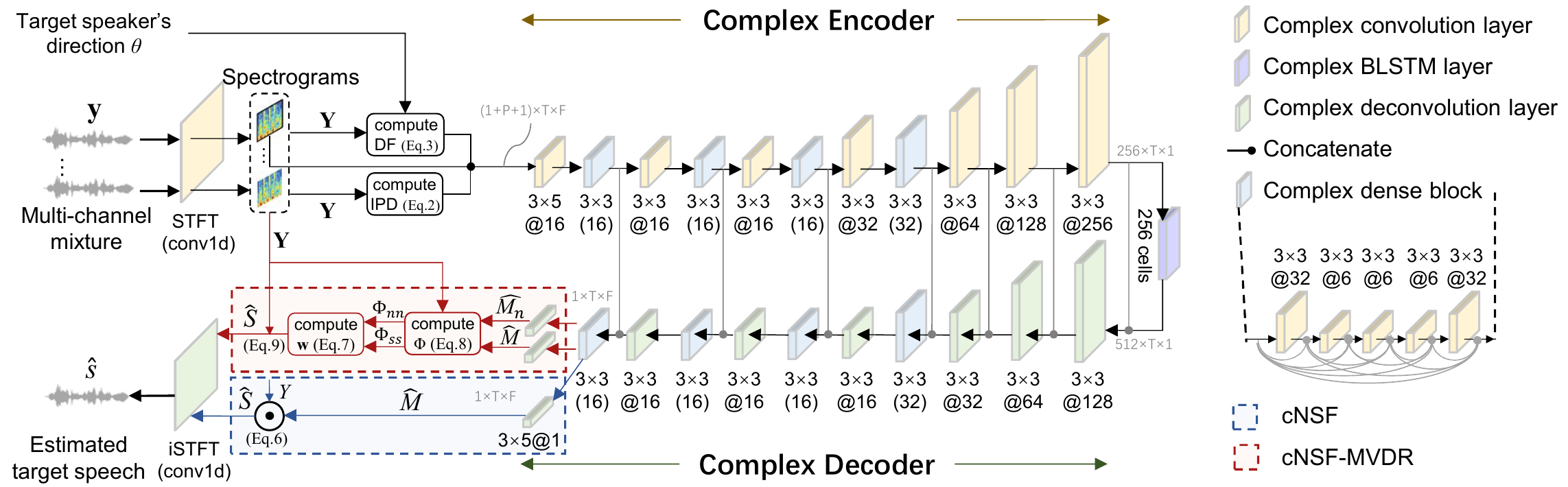}}
\vspace{-0.2cm}
\caption{The network architecture of our proposed complex neural spatial filter (cNSF) TSS model, where $\mathbf{y}$ is the multi-channel mixture, $\hat{M}$ and $\hat{M}_n$ are respectively the estimated cRM for target speech and remaining noise, $\hat{S}$ is the estimated complex spectrogram of target speech.}
\label{fig:net}
\end{figure*}

In this work, TSS aims to separate the target speech $s$ from the $U$-channel mixture $\mathbf{y}$ with given direction $\theta$ of the target speaker. We assume that the oracle direction of the target speaker is known by the separation system, via the camera or speaker localization frontends. Figure \ref{fig:net} illustrates our proposed TSS framework. The framework consists of three main modules: 1) feature extraction module (Sec \ref{subsec:trainingpair}), which leverages the temporal-spectral-spatial information in the multi-channel mixture signal to extract effective cues for TSS; 2) the backbone network (Sec \ref{subsec:net}) designed in the form of U-Net, including a complex encoder, complex BLSTM layers and a complex decoder; 3) target speech reconstruction module (Sec \ref{subsec:rec}). This module has two optional paths, one is the cRM based speech reconstruction (cNSF) and another is the cRM based MVDR beamforming (cNSF-MVDR).

\vspace{-0.1cm}
\subsection{Training pair formulation}
\label{subsec:trainingpair}
Following the supervised learning paradigm, the training target is to estimate the cRM of the target speech at the reference channel (the first channel in this work) as Eq. \ref{eq:crm}. To fit the cRM target defined in complex domain, we reformulate all the input features into complex-valued representations. In order to fully exploit the temporal-spectral-spatial information, spectral, spatial and directional features are combined as input features, following our previous work neural spatial filter (NSF) \cite{gu2019neural}.
To be specific, firstly, the reference channel of mixture spectrogram $Y\in \mathbb{C}^{T\times F}$ is taken as the spectral feature, where $T$ and $F$ are the total number of frames and frequency bands, respectively. In our implementation, the STFT kernel is reformulated as a complex convolution kernel \cite{gu2019end} to support end-to-end training. Then, since we assume a far-field scenario, IPD is chosen as the spatial feature considering that spatial information contained in interaural level difference (ILD) is relatively weak and noisy. IPD is computed as the phase difference between two channels of the complex spectrogram:
\vspace{-0.1cm}
\begin{equation}
\text{IPD}^{(p)}_{t,f} = \angle \mathbf{Y}^{p_1}_{t,f} - \angle \mathbf{Y}^{p_2}_{t,f}
\label{eq:ipd}
\end{equation}
where $t$ is the frame index, $f$ denotes the frequency, $\mathbf{Y}$ is the complex spectrogram of multi-channel mixture, $p_1$ and $p_2$ are two microphones of the $p$-th microphone pair, $P$ is the number of selected microphone pairs. In the ideal anechoic environment, IPD can be approximated as a linear function of frequency with a certain time delay \cite{watanabe2017new}. Since T-F bins that dominated by the same directional source experience the same time delay, IPDs of these bins will form a cluster within each frequency band. This property enables IPDs to indicate the T-F bin groups that dominated by the same source. However, as Wang et al. \cite{wang2019combining} pointed out, due to the phase wrapping issue, IPDs are not continuous across frequencies, which brings difficulty for the model to learn the correlation across frequency. Moreover, the linear functions that belong to different sources may cross because of spatial aliasing. To alleviate the discontinuity and crossing problem, we regard IPD as a complex-valued representation, of which the real and imaginary components are cosIPD and sinIPD, respectively. The cosIPD along with sinIPD formulates a 3d helix-like function \cite{wang2019combining}, therefore reducing the crossing in 3d space while promoting continuity.

Finally, to leverage the direction information of the target speech, we design a directional feature (DF). The formulation of DF aims to indicate the target speech dominance against spatially distributed interferences at each T-F bin, which is a discriminative feature for TSS \cite{chen2018multi, gu2019neural}. Specifically, define target-dependent phase difference $\text{TPD}_f^{(p)}(\theta)= 2\pi f \tau^{(p)}(\theta)$ \cite{gu2020multi} as the phase delay for the plane wave from $\theta$, with frequency $f$, to traverse between the $p$-th pair of microphones, where $\tau^{(p)}(\theta)$ is the corresponding pure delay. The design principle of DF is that, if a T-F bin is dominated by the target speech (from $\theta$), the modulus of Hermitian inner product between $\text{IPD}^{(p)}$ and $\theta$ related $\text{TPD}^{(p)}$ of this T-F bin will be close to 1, otherwise it will be 0. Following this concept, DF is computed as follows:
\vspace{-0.1cm}
\begin{equation}
d_{t,f}(\theta) = {\sum}^{P}_{p=1} \text{TPD}^{(p)}_{f}(\theta) \overline{\text{IPD}}^{(p)}_{t,f}
\label{eq:df}
\end{equation}
Since both TPD and IPD are complex, their inner product, i.e., the computed DF, is also complex-valued. From the above, all input features can then be obtained by concatenating them along the feature axis, i.e., $[Y, \text{IPD}^{(1)}, ..., \text{IPD}^{(P)}, d(\theta)]^{\mathsf{T}}$.

\vspace{-0.3cm}
\subsection{Network Design}
\label{subsec:net}

Borrowing design ideas from \cite{wang2020deep}, to achieve hierarchical feature abstraction, our network adopts a U-Net structure \cite{ronneberger2015unet}, which is featured with an encoder-decoder structure and skip connections. The convolutional and deconvolutional layers (yellow and green blocks in Figure \ref{fig:net}) are used to capture and retrieve the local information, respectively. The BLSTM layers added between the encoder and decoder are used to enhance the temporal sequence information modeling. Also, the densely connected layers (blue blocks) support efficient multi-level feature re-use \cite{huang2017densely}.

\begin{figure}
\centerline{\includegraphics[width=\linewidth]{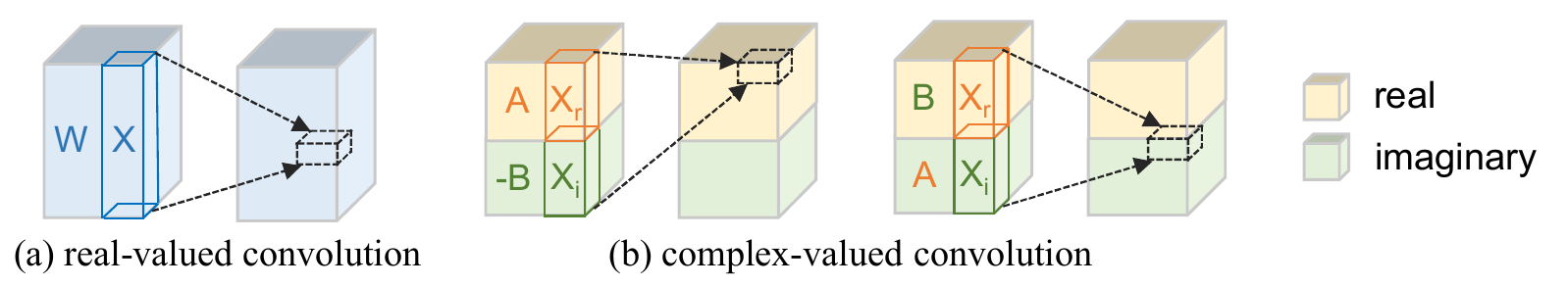}}
\vspace{-0.2cm}
\caption{Illustration of (a) real-valued convolution and (b) complex-valued convolution.}
\label{fig:cconv}
\end{figure}

In Figure \ref{fig:net}, the complex encoder is composed of alternating complex 2D convolutional (conv2d) layers and complex dense blocks. The complex convolution is defined between the complex input $X=X_r+jX_i$ and the complex kernel $W=A+jB$. As illustrated in Figure \ref{fig:cconv}(b), the real and imaginary parts of convolution operation can be written in matrix notation:
\vspace{-0.1cm}
\begin{equation}
\left[ \begin{array}{c}
{\left( W \circledast X\right)_r} \\ {\left( W \circledast X\right)_i}
\end{array}\right ] =
\left[ \begin{array}{cc}
A&{-B} \\ B&A
\end{array}\right ] \circledast
\left[ \begin{array}{c}
{X_r} \\ {X_i}
\end{array}\right ]
\label{eq:cconv}
\end{equation}
where $\circledast$ denotes the real-valued convolution operation. Each complex conv2d layer is specified in the format: \emph{height(time)}$\times$\emph{width(freq)}@\emph{featureMaps}, and the stride and padding is set as (1, 2) and (1, 0) for all layers. Each complex conv2d layer is followed by complex ReLU activation function along with the complex batch normalization \cite{trabelsi18deepcomplex}. Complex ReLU applies ReLUs on the real and the imaginary parts of a neuron separately. Complex batch normalization can be viewed as whitening 2-dimensional vectors, which makes the whitened complex input subject to the standard complex distribution with zero mean and unit covariance \cite{trabelsi18deepcomplex}. For dense blocks, each block (\emph{featureMaps}) contains 5 complex conv2d layers that are densely connected, and the stride and padding are set as (1, 1) and (1, 1). The complex BLSTM layer is implemented according to \cite{hu2020dccrn}, the output of which is given by:
\vspace{-0.05cm}
\begin{equation}
\begin{split}
    \text{cBLSTM}(X_r, X_i) &= \left( \text{BLSTM}_r(X_r) - \text{BLSTM}_i(X_i) \right) \\
&+ j\left( \text{BLSTM}_r(X_i) + \text{BLSTM}_i(X_r) \right)
\end{split}
\label{eq:cblstm}
\end{equation}
where $\text{cBLSTM}(\cdot)$ and $\text{BLSTM}(\cdot)$ represent the complex-valued and real-valued BLSTM processing, respectively. The complex decoder is composed of alternating complex 2D deconvolutional layers and complex dense blocks.
At the last deconvolutional layer, the number of output channel is set as 1 to predict the cRM of the target speech.

\vspace{-0.3cm}
\subsection{Target Speech Reconstruction}
\label{subsec:rec}

To reconstruct the target speech spectrogram based on the predicted cRM, we explore two ways. The first one (dashed blue box in Figure \ref{fig:net}), named as cNSF, obtaining the estimated target speech spectrogram $\hat{S}$ by directly multiplying the estimated complex mask $\hat{M}$ to $Y$:
\vspace{-0.1cm}
\begin{equation}
\hat{S}=\hat{M} \cdot Y = ( Y_r \hat{M_r} - Y_i \hat{M_i} ) + j ( Y_r \hat{M_i} + Y_i \hat{M_r} )
\label{eq:ests_m}
\end{equation}

\vspace{-0.1cm}
To reduce nonlinear speech distortions while further attenuating the interferences from the undesired direction, the cNSF-MVDR is developed (dashed red box in Figure \ref{fig:net}). MVDR is a classic and effective spatial beamformer proposed in \cite{capon1969high}, optimized with a constraint that minimizes the noise level without distorting the target speech. In our work, we follow the MVDR computation method in \cite{chen2008microphone}, and the coefficients $\mathbf{w}\in \mathbb{C}^{U \times F}$ can be computed as follows:
\begin{equation}
\mathbf{w}(f)=\frac{\Phi_{nn}^{-1}(f) \Phi_{ss}(f)}{\text{Tr} \left( \Phi_{nn}^{-1}(f) \Phi_{ss}(f)\right)}\mathbf{u}
\label{eq:wf}
\end{equation}
where $\mathbf{u}\in \mathbb{R}^{U}$ is a one-hot vector that marks the reference channel, $\text{Tr}(\cdot)$ solves the trace of a matrix. $\Phi_{ss}(f)$ and $\Phi_{nn}(f)$ are the spatial correlation matrices (SCM) of target speech and remaining noise, respectively. Motivated by the joint training of DNN and MVDR beamformer \cite{xu2020neural}, an MVDR neural beamformer is designed with the estimated cRM. Given the $\hat{M}$, the SCM of the target speech can be computed as:
\vspace{-0.1cm}
\begin{equation}
\Phi_{ss}(f) = \frac{\sum^{T-1}_{t=0} ( \hat{M}(t,f) \mathbf{Y}(t,f)) ( \hat{M}(t,f) \mathbf{Y}(t,f)) ^\mathsf{H}}{\sum^{T-1}_{t=0} ( \hat{M}^\mathsf{H}(t,f) \hat{M}(t,f))}
\label{eq:phi_ss}
\end{equation}
where $\hat{M}$ is shared across all the signal channels, $\mathsf{H}$ denotes the conjugate transpose matrix. In the same way, to compute $\Phi_{nn}(f)$, the cRM for remaining noise needs to be estimated as $\hat{M}_n$. Finally, the beamformed signal can be obtained by:
\vspace{-0.15cm}
\begin{equation}
\hat{S}(t,f)=\mathbf{w}^\mathsf{H}(f)\mathbf{Y}(t,f)
\label{eq:ests_mvdr}
\end{equation}

The iSTFT operation implemented as a complex 1D deconvolutional layer then converts the target complex spectrogram $\hat{S}$ to waveform $\hat{s}$.

\vspace{-0.3cm}
\subsection{Loss Function}
For the objective function, in order to optimize the whole network in the  end-to-end manner, we adopt the speech separation metric SI-SDR \cite{le2019sdr}. It aims to reduce the logarithm estimation error between estimated target speech $\hat{s}$ and ground truth $s$ in time domain, which is defined as $\text{SI-SDR}:=10\log_{10} ( \left\|s_{\text{target}}\right\|^{2} / \left\|e_{\text{noise}}\right\|^{2} )$, where $s_{\text{target}}:=(\left<\hat{s}, s\right>s) / \left\|s\right\|^{2}$, $e_{\text{noise}}:=\hat{s}-s_{\text{target}}$.

\section{Experiments and Results}
\label{sec:exp}

\subsection{Dataset}
To evaluate the proposed TSS framework under a challenging scenario, we simulate a multi-channel noisy reverberant dataset using the simulation pipelines described in \cite{gu2020multi}. The original speech data is collected from Youtube \cite{ailab2019large}, in which Mandarin accounts for the majority. The sampling rate is 16 kHz. The simulated dataset contains 192,000, 3,000 and 5,000 mixtures for training, validation and testing, respectively. The speakers appear in the test set has no overlap with the training set. The average duration of the testing utterance is about 4.1 seconds. We use a 15-element non-uniform linear array, with spacing 6-5-4-3-2-1-1-2-3-4-5-6 cm. The multi-channel speech signals are generated by convolving single-channel signals with room impulse responses (RIRs) simulated by image-source method \cite{ISM}. The room size is ranging from 4m-4m-2.5m to 10m-8m-6m (length-width-height). The reverberation time T60 is sampled in a range of 0.05s to 0.7s. The signal-to-interference ratio is ranging from -6 to 6 dB. Also, different types of noise are added with 18-30 dB signal-to-noise ratio.

\vspace{-0.3cm}
\subsection{Training and evaluation configuration}
For short time Fourier transform setting, we use 32ms square-root Hann window with 16ms hop size. IPDs and TPDs are extracted between 9 microphone pairs (1,15), (2,14), (3,13), (1,7), (12,4), (11,5), (12,8), (7,10), and (8,9) to sample different microphone spacings. Real-valued input features are the concatenation of $|Y|$, $\cos\text{IPD}$ and $|d(\theta)|$, while complex-valued input features are as described in Section \ref{subsec:trainingpair}.

The detailed hyper-parameters for cNSF is illustrated in Figure \ref{fig:net} with 17.5M parameters. As for NSF, to keep the same parameters with cNSF for fair comparison, the numbers of all the convolution channels and BLSTM cells are doubled. To evaluate the effectiveness of the proposed network structure (densely connected U-Net, DUNet), we also train a BLSTM based network for comparison \cite{gu2019neural}, in which there are four 256-cell BLSTM layers followed by two feedforward layers. Both the DUNet and BLSTM based network are trained with 4-second mixture chunks, using Ranger optimizer \cite{yong2020gradient} with early stopping. The learning rate is initialized as 1e-3 and will be decayed by 0.5 when the validation loss has no improvement for consecutive 3 epochs. \textit{pytorch\_complex} package is used to train the cNSF-MVDR.
Note that the multi-tap setup in \cite{xu2020neural} is not adopted to save the computation cost.

Following the evaluation pipeline in \cite{gu2020multi}, the performance is evaluated under different speaker mixing conditions: 1 speaker, 2 speakers and 3 speakers, respectively accounts for 5\%, 45\% and 50\% in the test set. All the speakers are assumed to not change their directions during speaking.
SI-SDR and WER are adopted as the evaluation metrics to measure the separation quality and speech intelligibility. The WER is measured using Tencent commercial Mandarin speech recognition API \cite{tencentapi}. The reverberant clean target speech is used as the reference for all the metric computations.

\begin{table}[t]
  \caption{SI-SDR (dB) and WER (\%) results of TSS systems. I denotes the input feature and O denotes the output target. }
  \label{tab:rlt}
  \vspace{-0.3cm}
    \small
\setlength{\tabcolsep}{3pt}
  \centering
  \begin{tabular}{l|c|c|p{20 pt}<{\centering}p{20 pt}<{\centering}p{20 pt}<{\centering}|p{20 pt}<{\centering}|c}
    \hline
     \hline
    \multirow{2}{*}{\textbf{Approach}} &
    \multirow{2}{*}{\textbf{I}} &
    \multirow{2}{*}{\textbf{O}} &
    \multicolumn{4}{c}{\textbf{SI-SDR (dB)}} &
    \multirow{2}{*}{\textbf{WER}} \\
    & & & 1spk & 2spk & 3spk & ave. \\
    \hline
    Mixture & - & - & 21.4 & 0.6 &-0.1 &1.3 &58.00\\
    Reverb. clean & - & - & $\infty$ & $\infty$ &$\infty$ & $\infty$ & 12.41\\
    \hline
     \hline
    NSF (BLSTM) & $\mathbb{R}$ & $\mathbb{R}$ & 23.3 & 10.4 & 8.8 & 10.2 & 29.10 \\
    NSF (BLSTM) & $\mathbb{R}$ & $\mathbb{C}$ & 23.4 & 10.5 & 8.9 & 10.3 & 28.88 \\
    NSF (BLSTM)  & $\mathbb{C}$ & $\mathbb{C}$ & 23.4 & 10.6 & 9.0 & 10.4 & 28.56 \\
    cNSF (BLSTM)  & $\mathbb{C}$ & $\mathbb{C}$ & 24.0 & 10.8 & 9.3 & 10.7 & 26.33 \\
    \hline
 NSF (DUNet)  & $\mathbb{R}$ & $\mathbb{R}$ & 23.4 &10.8 & 9.2 & 10.6 & 25.96 \\
 NSF (DUNet)  & $\mathbb{R}$ & $\mathbb{C}$ & 23.4 & 10.9 & 9.2 & 10.7 & 25.48 \\
 NSF (DUNet) & $\mathbb{C}$ & $\mathbb{C}$ & 23.7 & 11.3 & 9.7 & 11.1 & 22.46 \\
 cNSF (DUNet) & $\mathbb{C}$ & $\mathbb{C}$ & \textbf{24.3} & \textbf{11.9} & \textbf{10.3} & \textbf{11.7} & \textbf{21.10} \\

 \hline
  NSF-MVDR (DUNet) & $\mathbb{R}$ & $\mathbb{C}$ & 22.2 & 11.7 & 9.9 & 11.3 & 18.04 \\
 cNSF-MVDR (DUNet) & $\mathbb{C}$ & $\mathbb{C}$ &\textbf{23.5} & \textbf{12.3} & \textbf{10.5} & \textbf{12.0} & \textbf{17.03} \\
 \hline
 \hline
  \end{tabular}
\end{table}

\begin{figure}[hbt]
\centerline{\includegraphics[width=\linewidth]{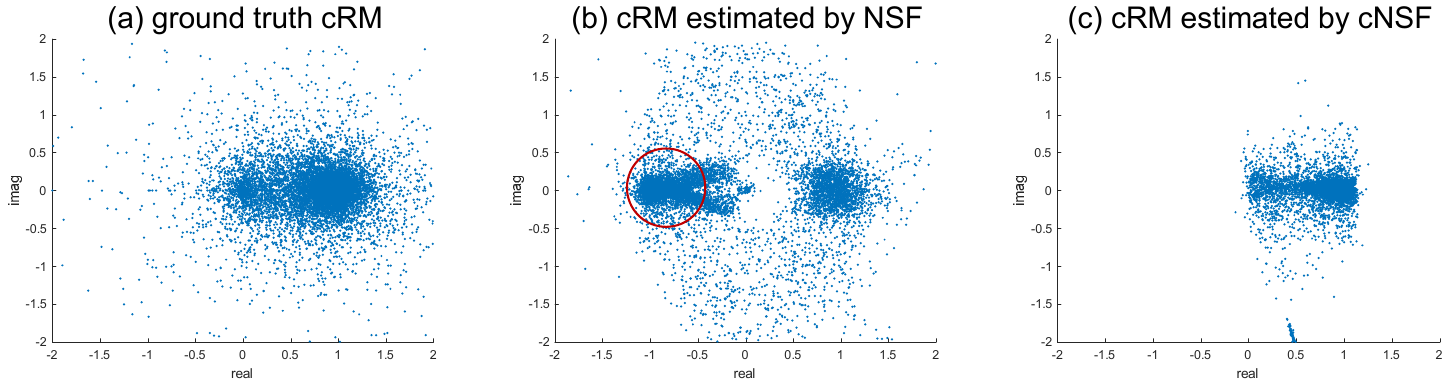}}
\vspace{-0.4cm}
\caption{The scatter plots of complex value distributions of the ground truth cRM, cRM estimated by NSF and cNSF on a 2-speaker mixture example. The red circle marks some complex values that form pairs of complex conjugates with corresponding ground truth cRM values. }
\label{fig:crm}
\end{figure}

\vspace{-0.5cm}
\subsection{Results analysis}
Table \ref{tab:rlt} reports the SI-SDR and WER results of the multi-channel TSS systems. The unprocessed mixture exhibits poor SI-SDR of 1.3 dB and WER of 58\%. The reverberant clean target speech, which is served as the reference for the metric computation, has a WER of 12.4\%.
Firstly, comparing results produced by BLSTM and DUNet based TSS models, it is obvious that DUNet based models exhibit consistent improvements over BLSTM based ones, owing to the effective multi-level feature abstraction ability of U-Net. Then, we examine the effects of training pair formulation on the separation performance. Beginning with the real-valued input features with RM target, DUNet based NSF obtains the SI-SDR of 10.6dB and WER of 26.0\%. Alternating the output target from RM to cRM, the SI-SDR has a gain of 0.1dB and WER reduces 0.5\%. Then, forming both input features and output target in complex domain using a real-valued DNN, the model further obtains a 0.4dB SI-SDR improvement and 3.0\% WER reduction. Finally, the proposed cNSF outperforms NSF by a large margin of 1.0dB SI-SDR and 4.4\% absolute WER.

When an MVDR module is used to reconstruct the target speech, performances of both NSF-MVDR and cNSF-MVDR improve over the NSF and cNSF, especially on the WER metric. This is because the target speech is comparatively preserved well, thanks to the distortionless constraint of MVDR formulation.
At last, cNSF-MVDR exhibits the best performance with SI-SDR of 12.0 dB and WER of 17.0\%.

To further examine the accuracy of the estimated cRM, Figure \ref{fig:crm} illustrates the distributions of cRMs on a two-speaker mixture. From the ground truth cRM plot (a), it is observed that most of the values are gathered around 0 and 1, while cRM estimated by NSF (b) misses the values around 0. Also, compared to the (a), the imaginary parts of some estimated complex values are opposite in sign (the red circle), which may lead to a right magnitude yet completely wrong phase. The cRM estimated by cNSF (c) shows a similar yet more compact distribution pattern with the ground truth, which confirms the accuracy of phase estimation and effectiveness of phase-aware processing with the proposed cDNN.

\vspace{-0.3cm}
\section{Conclusion}
\label{sec:conclusion}

In this work, we propose a novel multi-channel TSS framework, which directly models the cRM estimation in complex domain, and elaborately designed to fully exploit the temporal-spectral-spatial information. Experiments show that the target speech estimated by the proposed framework achieves better quality and intelligibility over the baseline.


\bibliographystyle{IEEEtran}
\bibliography{main}

\end{document}